\documentclass[aps,preprint]{revtex4}
\usepackage{graphicx,amssymb}
\begin{document}
\setcounter{page}{1}

\title{Temperature Dependence of a Width of $\Delta$H = $\Delta$B Region  \\ in 5 wt.\% (Fe, Ti) Paticle-Doped MgB$_2$  \\Superconductor}

\author{H. B. \surname{Lee}}
\author{G. C. \surname{Kim}}
\author{Y. C. \surname{Kim}}
\email{yckim@pusan.ac.kr}
\thanks{Fax: +82-51-513-7664}
\affiliation{Department of Physics, Pusan National University, Busan 46241, Korea}
\begin{abstract}

 A temperature dependence of a width of $\Delta$H = $\Delta$B region  has been studied for 5 wt.\% (Fe, Ti) particle-doped MgB$_2$ superconductor. The result  revealed  that widths of the region are linear along temperature. Here we show the meaning of the result and details of the calculation. In previous report, we represented a theory that a width of $\Delta$H = $\Delta$B region is related with upper critical field of the superconductor, which is that pinned fluxes at volume defect are picked out and move in $\Delta$H = $\Delta$B region when a distance between them is the same as that of upper critical field. Thus, we inspected the relationship between a width of the region and  upper critical field along temperature.  The theory would gain another justification if temperature dependence of a width of the region is proportional to that of upper critical field. 
We discussed several topics for $\Delta$H = $\Delta$B region of 5 wt.\% (Fe, Ti) particle-doped MgB$_2$ superconductor, which are Fe of (Fe, Ti) particle, Bean model, volume dependence of the region, etc..
 \end{abstract}



\maketitle

\section{Introduction} 
$\Delta$H = $\Delta$B region in a field dependence of magnetization curve (M-H curve) is the  region that appears in the volume defect-dominating superconductors.
 If  volume defects are many enough in the superconductor, $\Delta$H = $\Delta$B region appears, which is the region that the increased applied field is the same as the increased magnetic induction after the superconductor reached  the maximum diamagnetic property. The region appears because volume defects pin magnetic quantum fluxes preferentially over the entire specimen \cite{Lee4}. 

A volume defect pins much more quantum fluxes than planar or line defect does because its free energy density is low, which depends on its volume. If pinned fluxes at a volume defect are many, they bend like bow because of their repulsive force between them.  Thus,  fluxes have been not pinned at volume defects are pushed forward, and they would be pinned at other volume defect. Therefore, there are almost no unpinned fluxes in the superconductor if volume defects are many enough. Magnetic quantum fluxes are pinned at volume defects step by step from the surface to center of the superconductor as applied magnetic field increases. In this way, magnetic quantum fluxes in a superconductor are preferentially pinned at volume defects over entire specimen.

On the other hand, when  pinned fluxes are picked out from the volume defect by increase of magnetic quantum fluxes caused by applied magnetic field,  the entire pinned fluxes at a volume defect are picked out all at once instead of being leaked out from the volume defects one by one \cite{Gurp, Bonevich}.  We would use the term $'$pick out$'$ instead of depinning because they are picked out at once when pinned fluxes at a volume defect are depinned. It is considered that they are picked out at once by the tension of them.  

We have represented the two conditions that pinned fluxes are picked out from a volume defect \cite{Lee4}. One is when $\Delta$F$_{pickout}$ is more than  $\Delta$F$_{pinning}$ and the other is when a distance between of pinned fluxes at a volume defect is the same as that of upper critical field (H$_{c2}$). The former is  applicable  before H$_{c1}'$ which is defined as the applied magnetic field showing maximum diamagnetic property (the first field in $\Delta$H = $\Delta$B region), and the latter is applicable  in $\Delta$H = $\Delta$B region. The justification of the latter is due to the fact that the neighborhood of the volume defect would lose superconductivity if the distance between pinned fluxes is narrower than that of H$_{c2}$, thus the concept of flux pinning will not be established in the state.

A width of $\Delta$H = $\Delta$B region decreases as temperature increases. The behavior seems to be caused by the fact that the coherence length ($\xi$) of a superconductor increases as temperature increases, 
thus the number of fluxes pinned at a volume defect decreases as temperature increases. Therefore, a width of $\Delta$H = $\Delta$B region decreases as temperature increases. Thus, it is needed to check the relationship between  H$_{c2}$ 
and a width of $\Delta$H = $\Delta$B region as the temperature increases because H$_{c2}$ is $\Phi_0/2\pi \xi^2$, where $\Phi_0$ is magnetic quantum flux \cite{Poole}.

Many reviews do not believe the reality of $\Delta$H = $\Delta$B region and related theories. 
We think that the denial is mainly caused by the fact that $\Delta$H = $\Delta$B region is different from the general theory of superconductivity. If we had showed the relation between a width of $\Delta$H = $\Delta$B region  and temperature, the theory and our assertion would be understood more clearly. Because a width of  $\Delta$H = $\Delta$B region is related to H$_{c2}$ of the superconductor in the theory, 
we believe that  the reality of $\Delta$H = $\Delta$B region and  the theory would be proved more clearly if a width of the region depends on temperature as  H$_{c2}$ does. 

\section{results and discussion}
 \subsection{Decisions of the width of $\Delta$H = $\Delta$B region}
\begin{figure}
\vspace{2cm}
\begin{center}
\includegraphics*[width=11cm]{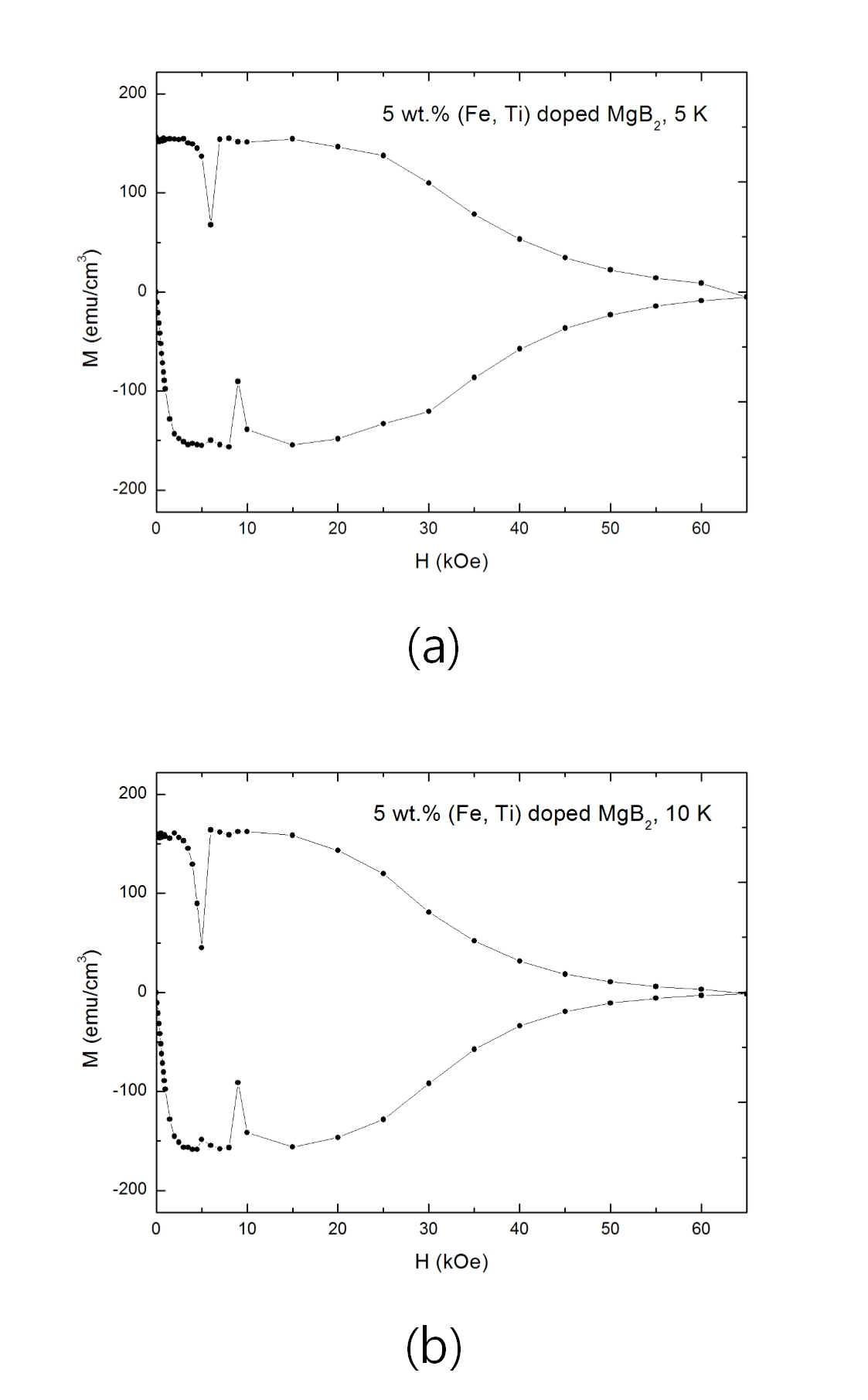}
\end{center}
\caption{ Field dependence of magnetizations (M-H curves) of 5 wt.\% (Fe, Ti) particle-doped MgB$_2$, which was air-cooled. (a): M-H curve at 5 K. (b): M-H curve at 10 K.}
\label{fig1}
\end{figure}

Widths of $\Delta$H = $\Delta$B region were obtained by taking average for the region of the first quadrant and that of the second quadrant in M-H curve. When the exact width is difficult to take due to flux jump, the value was reasonably taken. 
Figure \ref{fig1} (a) shows M-H curve of 5 wt.\% (Fe, Ti) particle-doped MgB$_2$ at 5 K. Inspecting the first and the second quadrants, it is certain that  widths of $\Delta$H = $\Delta$B region were expanded to 1.5 T. Thus the width is decided to be 1.3 T, which is 1.5 T- 0.2 T (H$_{c1}'$ is 0.2 T). 

M-H curve of the specimen at 10 K  is shown in Fig. 1 (b). 
The diamagnetic property at 1.5 T of 10 K is somewhat smaller than that of 5 K in first quadrant, and diamagnetic property of second quadrant at  1.5 T is slightly lower from the horizontal line. Thus, it is hard to think the region to be expanded to 1.5 T. Taking the median value, the width of the region is  1.05 T, which is 1.25 T - 0.2 T.

\begin{figure}
\vspace{2cm}
\begin{center}
\includegraphics*[width=11cm]{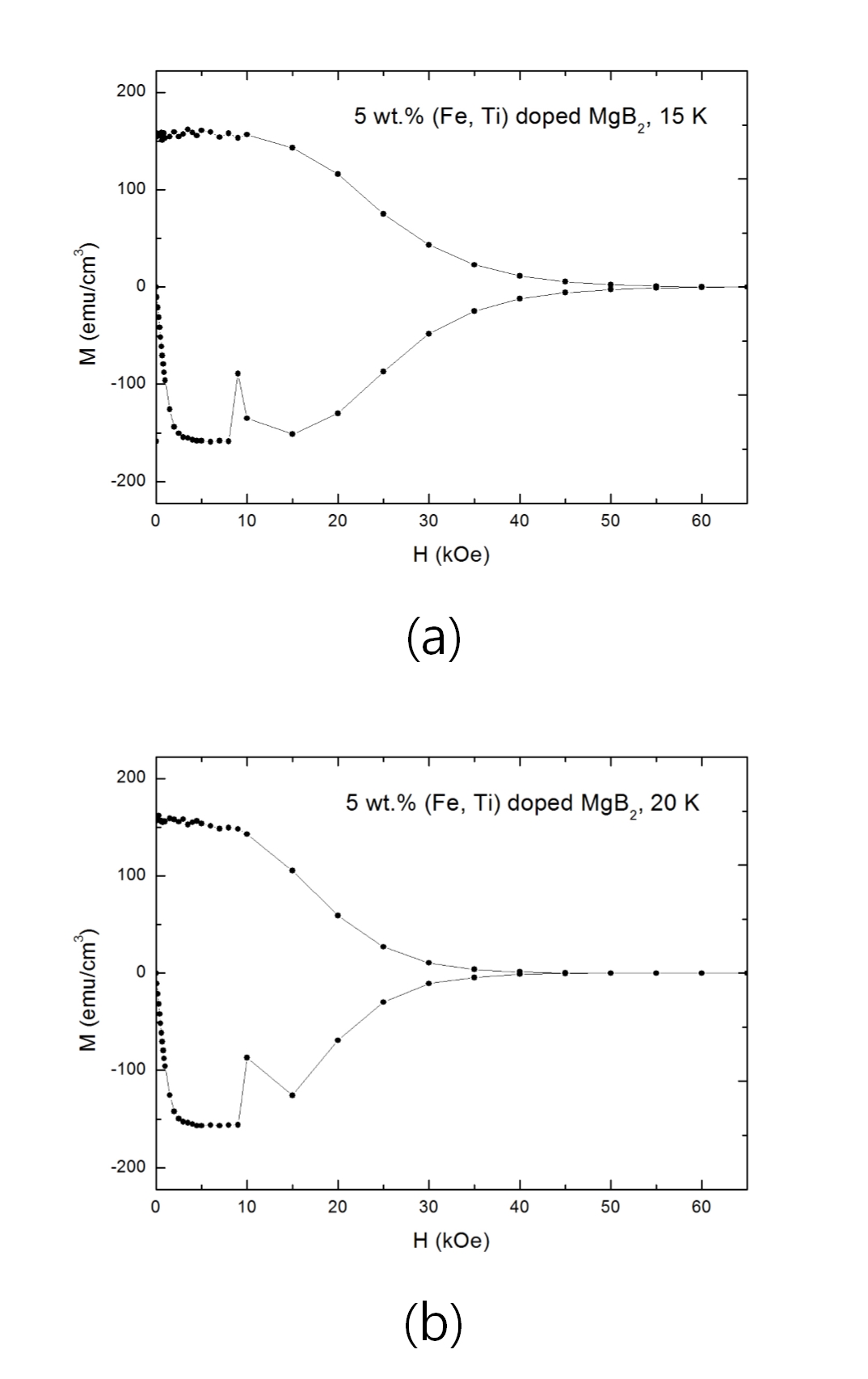}
\end{center}
\caption{ Field dependence of magnetizations (M-H curves) of 5 wt.\% (Fe, Ti) particle-doped MgB$_2$, which was air-cooled. (a): M-H curve at 15 K. (b): M-H curve at 20 K.}
\label{fig2}
\end{figure}

Figure \ref{fig2} (a) shows  M-H curve of the specimen at 15 K.  It is difficult to decide the exact value of the width in the first quadrant because of the flux jump. However,  it is certain that the region would be up to 1 T when considered in the second quadrant. Thus, the width at 15 K is 0.8 T, which is 1.0 T - 0.2 T.  M-H curve of the specimen at 20 K is shown in Fig. 2 (b). It is clear that the width of the region is up to 0.9 T in first quadrant. However, there is a little confusion that $\Delta$H = $\Delta$B region is up to 0.8 T or 0.9 T in second quadrant, thus the end of the region is conservatively decided to be 0.8 T. Therefore, the width of the region is 0.6 T, which is 0.8 T - 0.2 T. 

\begin{figure}
\vspace{2cm}
\begin{center}
\includegraphics*[width=11cm]{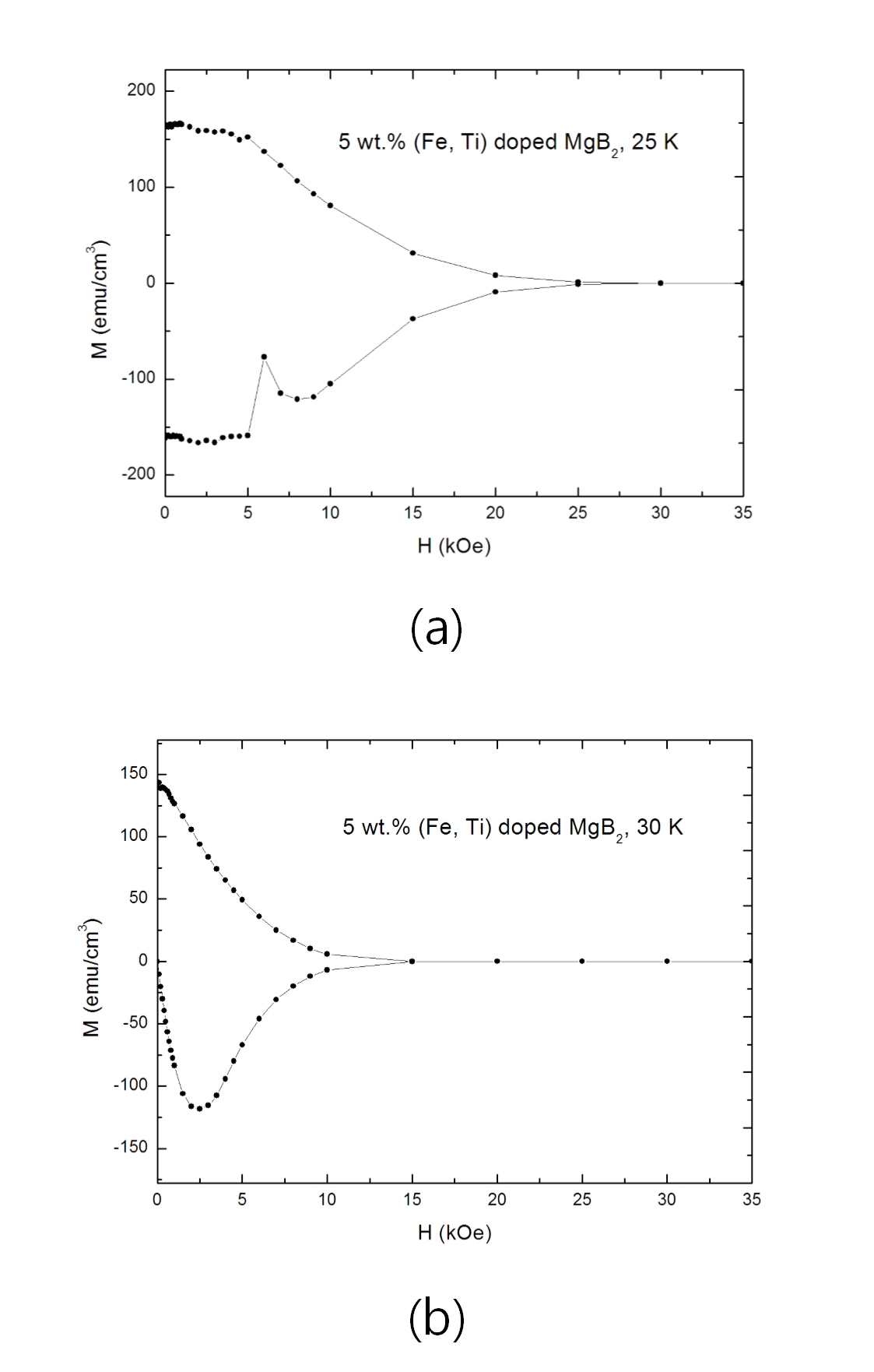}
\end{center}
\caption{ Field dependence of magnetizations (M-H curves) of 5 wt.\% (Fe, Ti) particle-doped MgB$_2$, which was air-cooled. (a): M-H curve at 25 K.  (b) M-H curve at 30 K }
\label{fig3}
\end{figure}

The M-H curve at 25 K of the specimen is shown in Fig. \ref{fig3} (a). 
The curve was made a mistake in measuring the magnetization without removing the pinned magnetic fluxes in the specimen. Estimating H$_{c1}'$ at 25 K, it is certain that it is higher field than that of 20 K and lower field than that of 30 K. Thus, it will be reasonable that H$_{c1}'$ at 25 K is  0.2 T because H$_{c1}'$ at 30 K is not far from 0.2 T. 
The region has been formed before the flux jump appears if  compared widths of the first quadrant with that of the second quadrant. Thus, the width of region at 25 K is determined to be 0.4 T, which is 0.6 T - 0.2 T. 
M-H curve of the specimen at 30 K is shown is Fig. 3 (c). $\Delta$H = $\Delta$B region completely disappears. The noting thing is the maximum diamagnetic property of the specimen also decreases a little.

The thickness of the specimen used for the measurement was 0.25 cm. Thus, the width of the region as unit thickness have to be 4$\times$t (t is the thickness of the specimen). In addition, since applied magnetic field penetrates both sides of the specimen,  volume defects in the superconductor have pinned the fluxes for both side until applied magnetic field reach H$_{c1}'$. 
Therefore, a width of the region as unit  thickness is
\begin{eqnarray}
 W_{\Delta H=\Delta B} = n_dw + (n_d-1)H_{c1}' 
 \end{eqnarray}  
where $W_{\Delta H=\Delta B}$ is a width of the region as unit thickness, $w$ is experimentally obtained width of the region, $n_d$ is the number of specimen when the specimen of unit thickness was divided ($n_dt$=1). 
Results for the width of $\Delta$H = $\Delta$B region along temperature are shown in Fig. 4 (a).  
As temperature increases, it is certain that a width of $\Delta$H = $\Delta$B region would decrease linearly. 

\subsection{Relation between the width of  $\Delta$H = $\Delta$B region and H$_{c2}$}
 A width of $\Delta$H = $\Delta$B region was suggested as follows \cite{Lee4}.
 \begin{eqnarray}
W_{\Delta H = \Delta B} = H_f - H_{c1}' = n^2m_{cps}m\Phi_0 - 4\pi M -  H_{c1}'
 \end{eqnarray} 
where $H_{c1}'$ is the first field of $\Delta$H = $\Delta$B region 
and $H_f$ is the final field of $\Delta$H = $\Delta$B region. $n^2$, m$_{cps}$, $m$, and $\Phi_0$ are the number of quantum fluxes pinned at a defect,  the number of defects which are in the vertically closed packed state, the number of defects having pinned fluxes from the surface to the center of the superconductor in an axis, and flux quantum, respectively.  m$_{cps}$ is the minimum number of defects when the penetrated fluxes into the superconductor are completely pinned. Thus, $2r$$\times$m$_{cps}$ is unit if $r$ is radius of spherical defect and $n^2m_{cps}m\Phi_0$ is magnetic induction in $\Delta$H = $\Delta$B region \cite{Lee4}.

The number of flux quanta that are pinned at a spherical defect of radius r 
 in a static state is
\begin{eqnarray}
n^2 = \frac{\pi r^2}{\pi (\frac{d}{2})^2}\times P = (\frac{2r}{d})^2\times P
 \end{eqnarray} 
 where $r$, $d$ and $P$ is a radius of defect, a distance between quantum fluxes pinned at the defect and filling rate which is $\pi$/4 when they have square structure, respectively \cite{Lee5}. 

If $d^2$ is 2$\pi$$\xi^2$, which is H$_{c2}$ state, a volume defect pin the maximum flux quanta. In addition, H$_{c2}$ is 
\begin{eqnarray}
 H_{c2} = \Phi_0/2\pi \xi^2 =\Phi_0/d^2
 \end{eqnarray} 
Inserting to Eq. (3)
\begin{eqnarray}
 n^2 = 
\frac{\pi r^2}{\Phi_0}H_{c2}
 \end{eqnarray} 
Thus,
 \begin{eqnarray}
W_{\Delta H = \Delta B} = \frac{\pi r }{2}mH_{c2} - 4\pi M -  H_{c1}'=aH_{c2}-b
 \end{eqnarray} 
Thus, $W_{\Delta H = \Delta B}$  is proportional to H$_{c2}$ if $r$ and $m$ are fixed. 
 Therefore, it is determined that the theory would have another justification if the behavior of the width along temperature is proportional to that of H$_{c2}$. 

 \subsection{Discussion}
It is clear that 5 wt.\% (Fe, Ti) particle-doped MgB$_2$ has $\Delta$H = $\Delta$B region in M-H curve. However, it is another problem that $\Delta$H = $\Delta$B region can be generalized in the superconductor. We insisted that the region can be generalized in the volume defect-dominating superconductor. However, many peer reviews thought that the region cannot  be generalized. We suggest two papers in literature, which concerned melt-textured growth superconductor that is one of volume defect-dominating superconductor.  $\Delta$H = $\Delta$B region in M-H curve was clearly shown in the papers \cite{Hsu, Muller}.

\begin{figure}
\vspace{2cm}
\begin{center}
\includegraphics*[width=11cm]{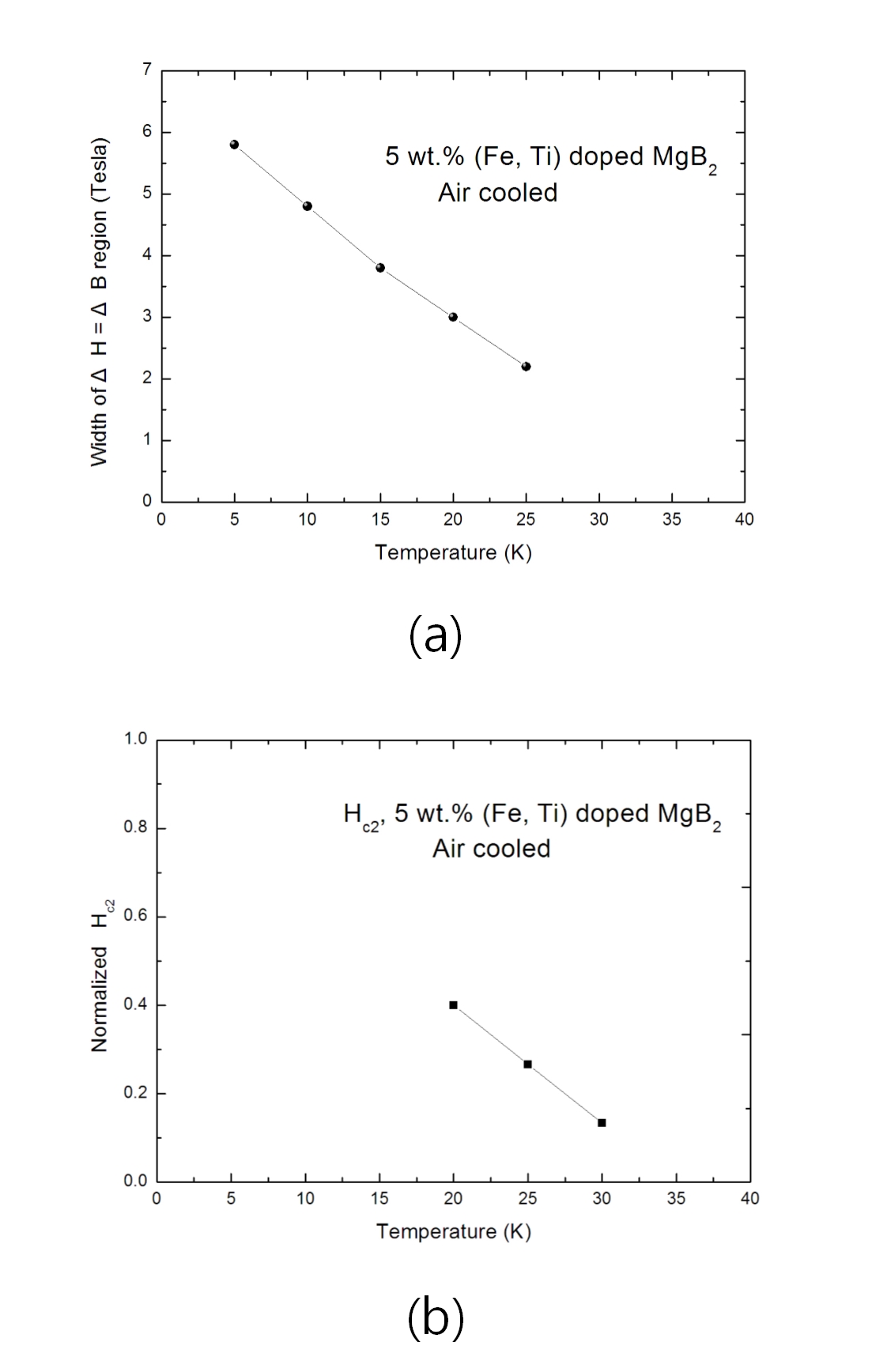}
\end{center}
\caption{ (a): Temperature dependence of a width of $\Delta$H =$\Delta$B region for 5 wt.\% (Fe, Ti) particle-doped MgB$_2$ (b):  Temperature dependence of upper critical field (H$_{c2}$) for 5 wt.\% (Fe, Ti) particle-doped MgB$_2$}
\label{fig4}
\end{figure}

The typical equation of H$_{c2}$  along temperature is as follows. 
\begin{eqnarray}
\xi(T)^2 
\propto\frac{1}{1-t}\Rightarrow
H_{c2}=\frac{\Phi_o}{2\pi\xi^2} \propto1-t
 \end{eqnarray} 
 where t = $T_m$/$T_c$, $\xi$ is  coherence length \cite{Tinkham}. $T_m$ is the measuring temperature  and  $T_c$ is critical temperature. The equation shows that H$_{c2}$ is linear along temperature. 
Fig. \ref{fig4} (b) is temperature dependence of H$_{c2}$ for  5 wt.\% (Fe, Ti) particle-doped MgB$_2$ specimen, which was air-cooled. It shows that H$_{c2}$ is also linear along temperature. Most of  literatures for MgB$_2$ were that H$_{c2}$ is linear along temperature \cite{Buzea, Eisterer}. Thus, it is clear that  H$_{c2}$ is linear along temperature. 

Therefore, it is certain that the pick-out of pinned fluxes at a volume defect is influenced by H$_{c2}$ because the temperature dependence of a width of $\Delta$H = $\Delta$B region is proportional to that of H$_{c2}$. On the other hand, we could understand the way that pinned fluxes are picked out from a defect in  $\Delta$H = $\Delta$B region if calculating $\Delta$F$_{pinning}$ and $\Delta$F$_{pickout}$. 
 As the result of calculating $\Delta$F$_{pinning}$ and $\Delta$F$_{pickout}$ in $\Delta$H = $\Delta$B region for 163 nm volume defects, $\Delta$F$_{pinning}$/$\Delta$F$_{pickout}$ was more than 4 \cite{Lee6}. Even though $\Delta$F$_{pinning}$ is more than four times larger than $\Delta$F$_{pickout}$, pinned fluxes at the volume defect have been picked out and moved in $\Delta$H = $\Delta$B region.  This is completely different from the concept that pinned fluxes are picked out and move when $\Delta$F$_{pickout}$ is more than $\Delta$F$_{pinning}$.

A noting thing is that a width of $\Delta$H = $\Delta$B region is dependent on a volume of the specimen, whereas  H$_{c2}$ is not. As applied magnetic field (H) increases,  fluxes are quantized and penetrate into the inside of the superconductor. If H approaches H$_{c2}$,  fluxes difference between inside and outside of superconductor is infinitesimal. Thus, H$_{c2}$ is not dependent of volume of the specimen. However, a width of the region decreases as much as a volume of a specimen decreases because the number of volume defects decreases as many as a volume of a specimen decreases. 



\begin{figure}
\vspace{2cm}
\begin{center}
\includegraphics*[width=11cm]{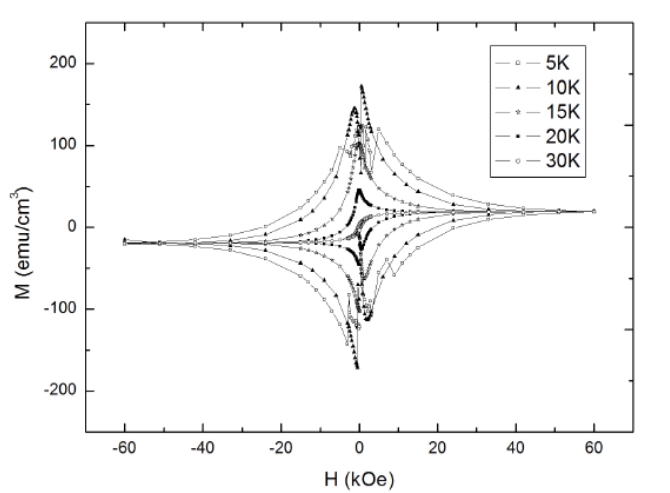}
\end{center}
\caption{ Distorted M-H curves of Fe-doped MgB$_2$ for various temperatures, Fe is nano-crystal particle.}
\label{fig5}
\end{figure}

Concerning Fe of (Fe, Ti) particle,  $\Delta$H = $\Delta$B region was suspected to be that M-H curve is distorted because of the ferromagnetic nature of Fe. This means that it was a mistake to assert that  $\Delta$H = $\Delta$B region appears in M-H curve of the specimen.  We already have known the problem of Fe, which is shown in Fig. \ref{fig5}.  It is certain that M-H curves were distorted if crystal Fe was doped on MgB$_2$. Especially, the specimen was broken after synthesizing MgB$_2$ because MgB$_2$ does not match crystal Fe. The figure was measuring a broken specimen. However, there was no problem with amorphous metal. (Fe, Ti) particle is amorphous, which shows different magnetic behavior from crystal Fe. We did not find any distortion in M-H curves of specimens 
\cite{Lee2}. Therefore, $\Delta$H = $\Delta$B region is real in the specimen as shown in Fig. 1 - 3.

Concerning  Bean model,  the flux penetration method represented in the previous paper violates  Bean model which is one of critical state model \cite{Lee4, bean}. We think that Bean model does not describe the nature of superconductor properly although it is honored. Researchers having studied the difference between the critical current density (J$_c$) of transport and that of Bean model would notice that the former is always larger than the latter. If Bean model was to describe the nature of superconductivity properly, the transport J$_c$ must have been less than that of Bean model because there are contact resistance, etc. Bean model seems to be a theory that suggests a minimum J$_c$ of the superconductor. In literature, it is observed that Fl$\ddot u$kiger et al. reported  transport J$_c$ and Bean model$'$s J$_c$ concurrently in Fe/MgB$_2$ tapes \cite{Flukiger}.

\section{Conclusion}
We studied temperature dependence of a width of $\Delta$H = $\Delta$B region in 5 wt. \% (Fe, Ti) particle-doped MgB$_2$ superconductor. A width of the region decreases linearly as temperature increases. In previous report, we asserted the theory that a width of $\Delta$H = $\Delta$B region relates with H$_{c2}$ and pinned fluxes are picked out from the volume defects when the distance between pinned fluxes is the same as that of H$_{c2}$. H$_{c2}$ of MgB$_2$ also decreases linearly as temperature increases. Therefore it is clear that a width of $\Delta$H = $\Delta$B region have relation with H$_{c2}$ of MgB$_2$. It is considered that the theory has another justification.

\section{Method}
 5 wt.\% (Fe, Ti) particle-doped MgB$_{2}$ and 5 wt.\% nano-Fe particle-doped MgB$_{2}$ were synthesized using the nonspecial atmosphere synthesis (NAS) method \cite{Lee}. 
 Briefing NAS method, Mg (99.9\% powder), B (96.6\% amorphous powder), (Fe, Ti) particles, and crystal nano-Fe particles were prepared for the specimen. After mixing Mg and B stoichiometry, (Fe, Ti) particles and crystal nano-Fe particles were added by weight, respectively. They were finely ground and pressed into 10 mm diameter pellets.  (Fe, Ti) particles were ball-milled for several days, and average radius of the (Fe, Ti) particles was approximately 0.163 $\mu$m. 
   On the other hand, an 8 m-long stainless-steel (304) tube was cut into 10 cm pieces. One side of the 10 cm-long tube was forged and welded. The pellets and excess Mg were placed in the stainless-steel tube. The pellets were annealed at 300$^o$C  for 1 hour to make them hard before inserting them into the stainless-steel tube. The other side of the stainless-steel tube was also forged. High-purity Ar gas was put into the stainless-steel tube, and which was then welded. The specimens was synthesized at 920$^o$C  for 1 hour and cooled in air. The field  dependences of magnetization were measured using a MPMS-7 (Quantum Design).

\end{document}